\documentclass[10pt,a4paper]{article}
\usepackage[utf8]{inputenc}
\usepackage[english]{babel}
\usepackage{amsmath}
\usepackage{graphicx}
\usepackage{amsfonts}
\usepackage{amssymb}
\usepackage{geometry}
\usepackage{float}
\usepackage{bm}
\usepackage{color}
\usepackage{cite}

\begin{document}

\title{Dunkl-Graphene in constant magnetic field}
\author{B. Hamil\thanks{%
hamilbilel@gmail.com} \\
D\'{e}partement de TC de SNV, Universit\'{e} Hassiba Benbouali, Chlef,
Algeria. 
\and B. C. L\"{u}tf\"{u}o\u{g}lu\thanks{%
bekircanlutfuoglu@gmail.com} \\
Department of Physics, University of Hradec Kr\'{a}lov\'{e}, \\
Rokitansk\'{e}ho 62, 500 03 Hradec Kr\'{a}lov\'{e}, Czechia.} 

\date{}
\maketitle

\begin{abstract}
Graphene-based materials are  thought to revolutionize entire industries. Therefore, many research are being carried on graphene theoretically and experimentally. On the other hand, recent studies show that the use of Dunkl derivative, instead of ordinary derivative, allows the concept of parity to be interpreted together with other physical quantities. In this manuscript, we investigate the thermal quantities of graphene under the constant magnetic field with the Dunkl-formalism.  We observe that only at low temperatures   Dunkl-parameters, thus parity, modify the conventional results.
\end{abstract}

\section{Introduction }

In the middle of the last century, Wigner thought of a very interesting question: "Do equations of motion determine quantum mechanical commutation relations?". To answer this question, he examined the harmonic oscillator problem and showed that the Heisenberg commutation relations may not always hold \cite{Wigner}. A year after Wigner's article, Born motivated Yang to investigate the problem within the context of wave mechanics \cite{Yang}. Yang realized that it is possible to hold the commutation relation if one uses the wave function conditions properly, including a more rigorous definition of Hilbert space and a more rigorous series expansion. To this end, he proposed a deformed Heisenberg algebra 
\begin{eqnarray}
\big[\hat{x},\hat{p}\big]=i\big(1+ \alpha \hat{R} \big),
\end{eqnarray}
by considering a reflection operator, $\hat{R}$,  with the given property $\hat{R} \, f(x) = f(-x)$, inside the deformed momentum operator 
\begin{eqnarray}
\hat{p}= i \bigg( -\frac{d}{dx}+ \frac{\alpha}{2x} \hat{R} \bigg).
\end{eqnarray}
Here, $\alpha$ denotes the Wigner parameter. Several decades later, Watanabe employed those modified operators to examine the quantum harmonic oscillator problem. He obtained two significant results:  the Hamiltonian operator has self-adjoint extensions for any Wigner parameter value, and the ground-state energy of the system is linearly associated to the Wigner parameter \cite{Watanabe}. 

In the same years, a discussion on the correlation of differential-difference and reflection operators from a purely mathematical point of view was going on \cite{Dunkl0}. In 1989, Charles F. Dunkl introduced a new combination of differential-difference and reflection operators, namely Dunkl operator, in the form of \cite{Dunkl1}: 
\begin{equation}\label{dunkloriginal}
D^{\mu_i}_{i}=\frac{\partial}{\partial x_i}+\frac{\mu_{i}}{x_{i}}(1-R),  \quad \quad {i=1,2,3.} \end{equation}
Dunkl operator fronted great developments in mathematics,  as one can expect \cite{Tsujimoto}. Surprisingly, physicists employed Dunkl-operator in several physical problems and Calogero-Sutherland-Moser  models \cite{Rosler, Lapointe, Kakei, vanDiejen, Polychronako, Plyushchay, Klishevich, Horvathy, Bie}. Especially, in the last decade, we observe an increasing interest in the Dunkl operator-based studies, since solutions provide additional information regarding the parity of the systems   \cite{G1, G2, G3, G4, G5, Ramirez1, Ramirez2, Sargol, Chung1, Ghaz, Mota1, Chungrev, Kim, Ojeda, Mota2, Mota3, Merad, ChungEPJP, Hassan, ChungDunkl, Dong, Bilel1, Bilel2, Mota2022}. 

In 2010, Geim and Novoselov were awarded the Nobel Price in Physics "for groundbreaking experiment regarding the two-dimensional material graphene" \cite{38}. In fact, awareness of graphene (graphite oxide) dates back to earlier than expected, to the mid-nineteenth century \cite{Brodie}. Since then, there are much research on the topic. For example, in 1947 Wallace predicted the electronic structure of graphite  \cite{Wallace}. In 1961, Boehm et al. obtained the thinnest sheets of graphite oxide, and then determined it with transmission electron microscopy \cite{Boehm}. Since 1987, a single layer of graphite has been referred to as graphene in literature \cite{Mouras}. In 2004, Geim and Novoselov with their collaborators used a simple mechanical method, a scotch tape, to extract thin layers of graphite crystal and identified the graphene with Atomic Force Microscopy \cite{Novoselov2004}. A graphene material has a two-dimensional structure with one-carbon atom thickness, where the atoms are lined up in a honeycomb lattice form \cite{40,39}. A graphene layer is the lightest but the strongest material, that presents superior mechanical, electronic and optical  features \cite{43, elastic, 41, Polin}. Thermal properties of graphene materials are extensively studied \cite{Balandin, Pop}. Graphene and its thermodynamics are also investigated in other phase-spaces, such as noncommutative  \cite{Bastos, Santos}, and curved spacetime with constant, negative and positive curvature \cite{Bilel, Gallerati, Iorio, Gallerati2}. 

Since graphene is the subject of extensive theoretical and experimental research of the last decade, we find interesting to revisit the problem in the Dunkl-formalism to interpret the  effect of parity on the thermal properties. Based on this motivation, we first construct the Dunkl-graphene Hamiltonian in the presence of constant magnetic field in Sec. 2. Then, in Sec. 3 we obtain the Dunkl-partition function and employ it to derive the Dunkl-Helmholtz free energy, Dunkl-internal energy, Dunkl-entropy and Dunkl-specific heat functions. After the illustration of the thermal quantities, we conclude the manuscript with a brief conclusion.

\section{Dunkl-Graphene in an external constant magnetic field}

In the theoretical formalism, a low energy excitation in a single graphene layer is described by the massless Dirac equation [45, 46].
\begin{equation}
\hat{H}\psi =E\psi ,
\end{equation}%
where 
\begin{equation}
\psi =\left( 
\begin{array}{c} 
\psi _{K} \\ 
\psi _{K^{\prime }}%
\end{array}%
\right) ,
\end{equation}%
which describes the electron states around each Dirac points $K$ and $K^{\prime }$  in the graphene case. Here, $\psi _{K},$ and $\psi _{K^{\prime }}$ are two dimensional eigenstates,%
\begin{equation}
\psi ^{K}=\left( 
\begin{array}{c}
\psi ^{A} \\ 
\psi ^{B}%
\end{array}%
\right), \quad  \quad \psi ^{K^{\prime }}=\left( 
\begin{array}{c}
\psi ^{A^{\prime }} \\ 
\psi ^{B^{\prime }}%
\end{array}%
\right) .
\end{equation}%
We start by expressing the Dunkl-Dirac Hamiltonian using the Dunkl derivative instead of the ordinary partial derivative
\begin{equation}
\hat{H}=\frac{v_{F}\hbar }{i}\overrightarrow{\alpha }\cdot \overrightarrow{D}.
\end{equation}%
Here, $\overrightarrow{\alpha }$ and $v_F$ denote the Dirac matrices and  Fermi velocity, respectively, while $v_{F}=\left(
1.12\pm 0.02\right)\times 10^{6}\,\mathrm{m/s}$. Now,
let us consider an external homogeneous magnetic field, $\overrightarrow{B}=B%
\overrightarrow{e_{z}}$, where we assume $B>0$ and employ $A=\frac{B}{2}\left( 
-y,x\right) $ gauge. Then, the Dunkl-Dirac Hamiltonian for the two Dirac
points reads
\begin{equation}
\hat{H}=\left( 
\begin{array}{cc}
H_{K} & 0 \\ 
0 & H_{K^{\prime }}%
\end{array}%
\right) ,
\end{equation}
where%
\begin{equation}
H_{K}=\hbar v_{F}\left( 
\begin{array}{cc}
0 & \frac{1}{i}D_{x}-D_{y}+\frac{eB}{2\hbar c}\left( y+ix\right) \\ 
\frac{1}{i}D_{x}+D_{y}+\frac{eB}{2\hbar c}\left( y-ix\right) & 0%
\end{array}%
\right) ,
\end{equation}%
and%
\begin{equation}
H_{K^{\prime }}=\hbar v_{F}\left( 
\begin{array}{cc}
0 & \frac{1}{i}D_{x}+D_{y}+\frac{eB}{2\hbar c}\left( y-ix\right) \\ 
\frac{1}{i}D_{x}-D_{y}+\frac{eB}{2\hbar c}\left( y+ix\right) & 0%
\end{array}%
\right) .
\end{equation}%
Via the decoupling of Dunkl-Dirac Hamiltonian to $H_{K}$ and $H_{K^{\prime }}$, we examine solution of the eigenvalue equation $H_{K}\psi _{K}=E\psi _{K}$. Using this, we easily obtain the following algebraic equations:
\begin{eqnarray}
\left[ \frac{1}{i}D_{x}-D_{y}+\frac{1}{2l_{B}^{2}}\left( y+ix\right) \right]
\psi ^{B}&=&\frac{E}{\hbar v_{F}}\psi ^{A},  \label{A} \\
\left[ \frac{1}{i}D_{x}+D_{y}+\frac{1}{2l_{B}^{2}}\left( y-ix\right) \right]
\psi ^{A}&=&\frac{E}{\hbar v_{F}}\psi ^{B},  \label{B}
\end{eqnarray}
where $l_{B}^{2}\equiv\frac{\hbar c}{eB}$ is the magnetic length. Then, we
substitute Eq. (\ref{B}) in Eq. (\ref{A}) and employ the Dunkl derivatives given in
Eq. \eqref{dunkloriginal}. We find the following differential equation for the component $\psi
^{A}$ 
\begin{equation}
\left \{ \Delta _{D}-\frac{y^{2}+x^{2}}{4l_{B}^{4}}-\frac{i\left(
xD_{y}-yD_{x}\right) }{l_{B}^{2}}+\frac{1+\mu _{2}R_{2}+\mu _{1}R_{1}}{%
l_{B}^{2}}+\frac{E^{2}}{\hbar ^{2}v_{F}^{2}}\right \} \psi ^{A}=0,  \label{C}
\end{equation}
where the Dunkl-Laplacian operator is
\begin{equation}
\Delta _{D}=\Delta +\frac{2\mu _{1}}{x}\frac{\partial }{\partial x}+\frac{%
2\mu _{2}}{y}\frac{\partial }{\partial y}-\frac{\mu _{1}\left(
1-R_{1}\right) }{x^{2}}-\frac{\mu _{2}\left( 1-R_{2}\right) }{y^{2}}.
\end{equation}

\subsection{Polar coordinate solutions}
In order to solve Eq. \eqref{C},
we prefer to use the polar coordinates
\begin{equation}
x=r\cos \theta ;\text{ \ }y=r\sin \theta.
\end{equation}
It is worth noting that in this case the reflection operators have the action%
\begin{equation}
R_{1}f\left( r,\theta \right) =f\left( r,\pi -\theta \right), \quad \text{ \ and \
\ }
\quad 
R_{2}f\left( r,\theta \right) =f\left( r,-\theta \right) .
\end{equation}%
After the algebra we find that Eq. \eqref{C} turns to%
\begin{equation}
\left[ \frac{\partial ^{2}}{\partial r^{2}}+\frac{1+2\mu _{1}+2\mu _{2}}{r}%
\frac{\partial }{\partial r}-\frac{r^{2}}{4l_{B}^{4}}-\frac{2\mathcal{B}%
_{\theta }}{r^{2}}-\frac{\mathcal{J}_{\theta }}{l_{B}^{2}}+\frac{1+\mu
_{2}R_{2}+\mu _{1}R_{1}}{l_{B}^{2}}+\frac{E^{2}}{\hbar ^{2}v_{F}^{2}}\right]
\psi ^{A}=0, \label{eq17}
\end{equation}%
where the angular operators $\mathcal{B}_{\theta }$ and $\mathcal{J}_{\theta
}$ are defined as following, respectively: 
\begin{eqnarray}
\mathcal{B}_{\theta }&=&-\frac{\partial ^{2}}{2\partial \theta ^{2}}+\frac{\mu _{1}}{2r^{2}\cos ^{2}\theta }\left( 1-R_{1}\right) +\frac{\mu _{2}}{2r^{2}\sin ^{2}\theta }\left( 1-R_{2}\right) +%
\frac{\mu _{1}\tan \theta -\mu _{2}\cot \theta }{r^{2}}\frac{\partial }{%
\partial \theta }, \\
\mathcal{J}_{\theta }&=&i\left( \frac{\partial }{\partial \theta }+\mu
_{2}\cot \theta \left( 1-R_{2}\right) -\mu _{1}\tan \theta \left(
1-R_{1}\right) \right) .
\end{eqnarray}
In \cite{G1}, it is shown that the square of the angular operator $
\mathcal{J}_{\theta }$ is linked to the operator $\mathcal{B}_{\theta }$ as%
\begin{equation}
\mathcal{J}_{\theta }^{2}=2\mathcal{B}_{\theta }+2\mu _{1}\mu _{2}\left(
1-R_{1}R_{2}\right) .
\end{equation}
Since $R_{1}R_{2}$ operator commutes with the operator $\mathcal{J}_{\theta }$, we can employ the eigenfunction 
\begin{equation}
\mathcal{J}_{\theta }g_{\epsilon }\left( \theta \right) =\lambda _{\epsilon
}g_{\epsilon }\left( \theta \right) ,  \label{D}
\end{equation}
where $s_{1}$, $s_{2}$ are the eigenvalues of the reflection operators $R_{1}$ and $R_{2}$, respectively, and thus, $\epsilon =s_{1}s_{2}=\pm 1$. Then, we determine the eigenfunctions, $g_{\epsilon}\left( \theta \right)$, and the eigenvalues, $\lambda _{\epsilon }$, in two different cases:
\begin{itemize}
\item In the $\epsilon =1$ case, Eq. (\ref{D})  leads to 
\begin{equation}
g_{+}=a_{\ell }\, \mathbf{P}_{\ell }^{\left( \mu _{1}+1/2,\mu _{2}+1/2\right)
}\left( -2\cos \theta \right) \pm a_{\ell }^{\prime }\sin \theta \cos \theta \,
\mathbf{P}_{\ell -1}^{\left( \mu _{1}+1/2,\mu _{2}+1/2\right) }\left( -2\cos
\theta \right) ,
\end{equation}%
and%
\begin{equation}
\lambda _{+}=\pm 2\sqrt{\ell \left( \ell +\mu _{1}+\mu
_{2}\right) },\quad\quad \ell \in 
\mathbb{N}
^{\ast },  \label{lambdapl}
\end{equation}%
where $\mathbf{P}_{\ell }^{\left( \alpha ,\beta \right) }\left( x\right) $ are
the classical Jacobi polynomials, while $a_{\ell }$, and $a_{\ell }^{\prime }$ are constants.

\item In the $\epsilon =-1$ case, Eq. (\ref{D})  gives 
\begin{equation}
g_{-}=b_{\ell }\cos \theta \, \mathbf{P}_{\ell -1/2}^{\left( \mu _{1}+1/2,\mu
-1/2\right) }\left( -2\cos \theta \right) \pm b_{\ell }^{\prime }\sin \theta \, 
\mathbf{P}_{\ell -1/2}^{\left( \mu _{1}-1/2,\mu _{2}+1/2\right) }\left(
-2\cos \theta \right) ,
\end{equation}%
and
\begin{equation}
\lambda _{-}=\pm 2\sqrt{\left( \ell +\mu _{1}\right) \left( \ell
+\mu _{2}\right) }, \quad\quad \ell \in \left \{ 1/2,3/2,...\right \} . \label{lambdaneg}
\end{equation}
\end{itemize}
Now, we assume that $\psi ^{A}\equiv \Psi ^{s_{1},s_{2}}\left( r\right)
g_{\epsilon }\left( \theta \right) $. Then, Eq. \eqref{eq17} turns to
\begin{equation}
\left[ \frac{d^{2}}{dr^{2}}+\frac{1+2(\mu _{1}+\mu _{2})}{r}\frac{d}{dr}-%
\frac{r^{2}}{4l_{B}^{4}}-\frac{\lambda _{\epsilon }^{2}-2\mu _{1}\mu
_{2}\left( 1-\epsilon \right) }{r^{2}}-\frac{\lambda _{\epsilon }}{l_{B}^{2}}%
+\frac{1+\mu _{1}s_{1}+\mu _{2}s_{2}}{l_{B}^{2}}+\frac{E^{2}}{\hbar
^{2}v_{F}^{2}}\right] \Psi ^{s_{1},s_{2}}\left( r\right) =0.  \label{E}
\end{equation}%
We observe that the solution of this differential equation splits into four-parity sectors which can be labeled by the eigenvalues of the quantum numbers $\epsilon$, and $s_{1}$, $s_{2}$.

\subsubsection{The $\epsilon =1$ case with  $s_{1}=s_{2}=1$}

In order to solve Eq. (\ref{E}), we define a new variable, $\xi= \frac{r^2}{2\ell_B^2}$. Taking this case's parameter values into account, we obtain a form of Kummer's differential equation 
\begin{equation}
\left[ \xi \frac{d^{2}}{d\xi ^{2}}+\left( 1+\mu _{1}+\mu _{2}\right) \frac{d%
}{d\xi }-\frac{\xi }{4}-\frac{\lambda _{+}^{2}}{4\xi }+\frac{1-\lambda
_{+}+\mu _{1}+\mu _{2}}{2}+\frac{l_{B}^{2}E^{2}}{2\hbar ^{2}v_{F}^{2}}\right]
\Psi ^{+,+}=0,  \label{F}
\end{equation}%
that has solution of the form
\begin{equation}
\Psi ^{+,+}=\xi ^{\ell }e^{-\frac{\xi }{2}}\mathbf{F}\left( \frac{2\ell +\lambda _{+}}{2}-\frac{%
l_{B}^{2}E^{2}}{2\hbar ^{2}V_{F}^{2}},1+2\ell +\mu _{1}+\mu _{2};\xi \right),  \label{G}
\end{equation}%
where $\mathbf{F}\left( a,b,x\right) $ is the confluent hypergeometric
function. As $\xi \rightarrow \infty (r\rightarrow \infty) $, the radial wave function should tend to zero. This requirement can be fulfilled when we employ the condition
\begin{equation}
\frac{2\ell +\lambda _{+}}{2}-\frac{l_{B}^{2}E^{2}}{2\hbar ^{2}V_{F}^{2}}=-n,
\label{K}
\end{equation}%
which leads to the quantization of the energy eigenvalues%
\begin{equation}
E_{n,\ell }^{+,+}=\pm \frac{\hbar v_{F}}{l_{B}}\sqrt{2\left(n+\ell \pm\sqrt{\ell
\left( \ell +\mu _{1}+\mu _{2}\right) }\right)}, \label{Enpp}
\end{equation}%
for positive integers, $n$. We observe that energy spectrum depends to the quantum numbers $\left( n,\ell \right) $ and the Dunkl parameters $\mu _{j}$. This fact is based on the modification of Heisenberg algebra that arose as a result of the introduction of Dunkl derivative operators.  We note that when $\ell=0$, the energy spectrum loses its dependency on the Wigner parameter and reduces to the ordinary quantum mechanic \cite{Bastos, Santos}.
\begin{equation}
E_{n, 0}=\pm \frac{\hbar v_{F}}{l_{B}}\sqrt{2n}. \label{ordEnspec}
\end{equation}
On the other hand, in the limit, ($\mu _{1}\rightarrow 0$ and $\mu _{2}\rightarrow 0$) or ($\mu _{1}+\mu_2\rightarrow 0$), the Dunkl-energy spectrum turns 
to 
\begin{eqnarray}
E_{n, \ell}=\pm \frac{\hbar v_{F}}{l_{B}}\sqrt{2(n+\ell\pm |\ell| ) }, \label{firstsol}
\end{eqnarray}
that has two solutions. The first one is identical to Eq. \eqref{ordEnspec}, while the second one is
\begin{eqnarray}
E_{n,\ell}=\pm \frac{\hbar v_{F}}{l_{B}}\sqrt{2(n+2\ell ) }, \label{secsol}
\end{eqnarray}
since $\ell \in \mathbb{N}^{\ast }$ .

\subsubsection{The $\epsilon =1$ case with  $s_{1}=s_{2}=-1$}
In this case, Eq. (\ref{E}) reduces to%
\begin{equation}
\left[ \xi \frac{d^{2}}{d\xi ^{2}}+\left( 1+\mu _{1}+\mu _{2}\right) \frac{d%
}{d\xi }-\frac{\xi }{4}-\frac{\lambda _{+}^{2}}{4\xi }+\frac{1-\lambda
_{+}-\mu _{1}-\mu _{2}}{2}+\frac{l_{B}^{2}E^{2}}{2\hbar ^{2}v_{F}^{2}}\right]
\Psi ^{-,-}=0, \label{L}
\end{equation}%
which is similar to previous case. Therefore, its solution can be obtained in the same manner. After performing those steps, the eigensolution takes the form %
\begin{equation}
\Psi ^{-,-}=\xi ^{\ell }e^{-\frac{\xi }{2}}\mathbf{F}\left( \frac{\lambda
_{+}+2\mu _{2}+2\mu _{1}}{2}+\ell -\frac{l_{B}^{2}E^{2}}{2\hbar ^{2}V_{F}^{2}%
},1+2\ell +\mu _{1}+\mu _{2};\xi \right) .
\end{equation}%
which vanishes at $\xi \rightarrow \infty $ when $\frac{\lambda _{+}+2\mu _{2}+2\mu _{1}}{2}+\ell -\frac{%
l_{B}^{2}E^{2}}{2\hbar ^{2}V_{F}^{2}}=-n$. Therefore, the Dunkl-energy eigenvalue function appears as %
\begin{equation}
E_{n,\ell }^{-,-}=\pm \frac{\hbar v_{F}}{l_{B}}\sqrt{2\left( n + \ell +\mu
_{1}+\mu _{2} \pm \sqrt{\ell \left( \ell +\mu _{1}+\mu _{2}\right) }\right)}. \label{Enmm}
\end{equation}
Alike from the previous case, when $\ell=0$, the energy spectra depends on the Wigner parameter
\begin{equation}
E_{n, 0}=\pm \frac{\hbar v_{F}}{l_{B}}\sqrt{2(n+\mu_1+\mu_2)}. 
\end{equation}
On the other hand, for ($\mu _{1}\rightarrow 0$ and $\mu _{2}\rightarrow 0$) or ($\mu _{1}+\mu_2\rightarrow 0$), the Dunkl-energy spectrum turns 
to Eq. \eqref{firstsol}.

\subsubsection{The $\epsilon =-1$ case with  $s_{1}=1, s_{2}=-1$}
In this case, we substitute Eq. (\ref{lambdaneg}) into Eq. (\ref{E}). We find
\begin{equation}
\left[ \xi \frac{d^{2}}{d\xi ^{2}}+\left( 1+\mu _{1}+\mu _{2}\right) \frac{d%
}{d\xi }-\frac{\xi }{4}-\frac{\lambda _{-}^{2}-4\mu _{1}\mu _{2}}{4\xi }+%
\frac{1+\mu _{1}-\mu _{2}-\lambda _{-}}{2}+\frac{l_{B}^{2}E^{2}}{2\hbar
^{2}v_{F}^{2}}\right] \Psi ^{+,-}=0,
\end{equation}%
which has a solution of the form
\begin{equation}
\Psi ^{+,-}=\xi ^{\ell }e^{-\frac{\xi }{2}}\mathbf{F}\left( \frac{2\mu
_{2}+\lambda _{-}}{2}+\ell -\frac{l_{B}^{2}E^{2}}{2\hbar ^{2}V_{F}^{2}}%
,1+2\ell+\mu _{1}+\mu _{2} ;\xi \right) .
\end{equation}%
Using the boundary condition, we obtain the energy quantization as 
\begin{equation}
E_{n,\ell }^{+,-}=\pm \frac{\hbar v_{F}}{l_{B}}\sqrt{2\left(n+\ell +\mu _{2} \pm \sqrt{\left( \ell +\mu _{1}\right) \left( \ell +\mu _{2}\right) }\right)}. \label{Enpm}
\end{equation}
We note that in this case $\ell$ is a positive half integer. Therefore, we can only examine the particular case in which the Wigner parameters take ($\mu _{1}\rightarrow 0$ and $\mu _{2}\rightarrow 0$) or ($\mu _{1}+\mu_2\rightarrow 0$). For ($\mu _{1}\rightarrow 0$ and $\mu _{2}\rightarrow 0$), the Dunkl-energy spectrum turns to Eq. \eqref{firstsol}. For ($\mu _{1}+\mu_2\rightarrow 0$), we have
\begin{eqnarray}
E_{n,\ell}&=&\pm \frac{\hbar v_{F}}{l_{B}}\sqrt{2\left(n+\ell +\mu _{2} \pm \sqrt{ \ell^2 -\mu _{2}^2 }\right)}. \label{Enpm2}
\end{eqnarray}
In this case, in order to avoid complex values, the condition $\ell\geq \mu_2$  should be satisfied. 

\subsubsection{The $\epsilon =-1$ case with  $s_{1}=-1, s_{2}=1$}
\begin{figure*}[htb!]
\resizebox{\linewidth}{!}{\includegraphics{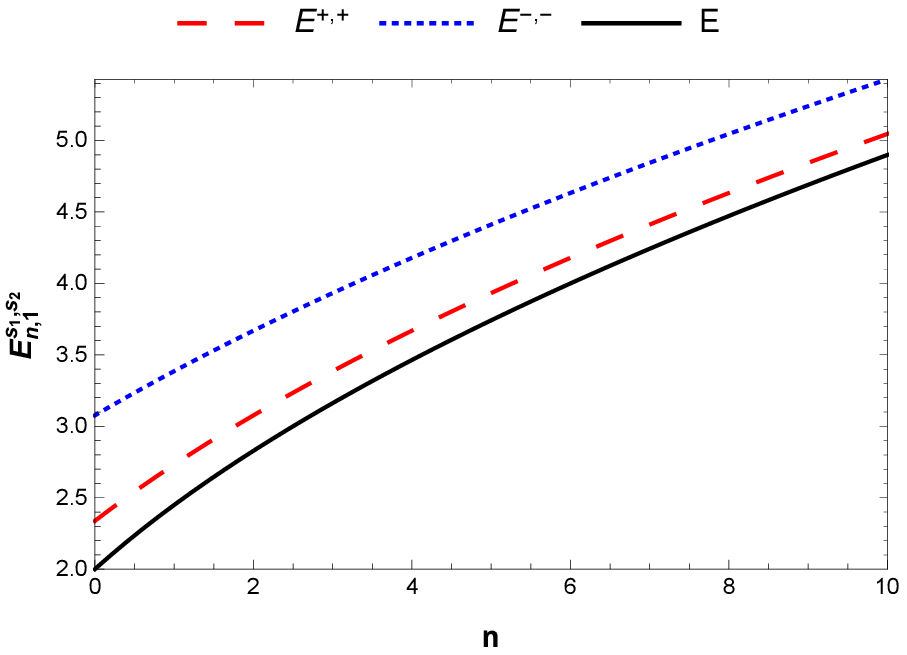},%
\includegraphics{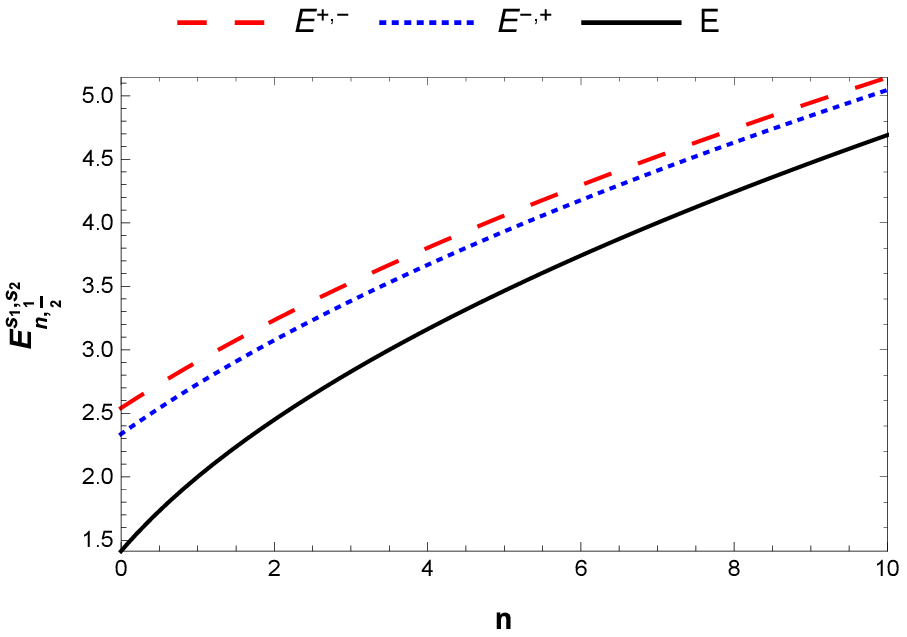}}
\caption{Dunkl-energy spectra of $\ell=1$ and $\ell=1/2$ cases, respectively, versus $n$.}\label{Fig1}
\end{figure*}

\begin{figure*}[htb!]
\resizebox{\linewidth}{!}{\includegraphics{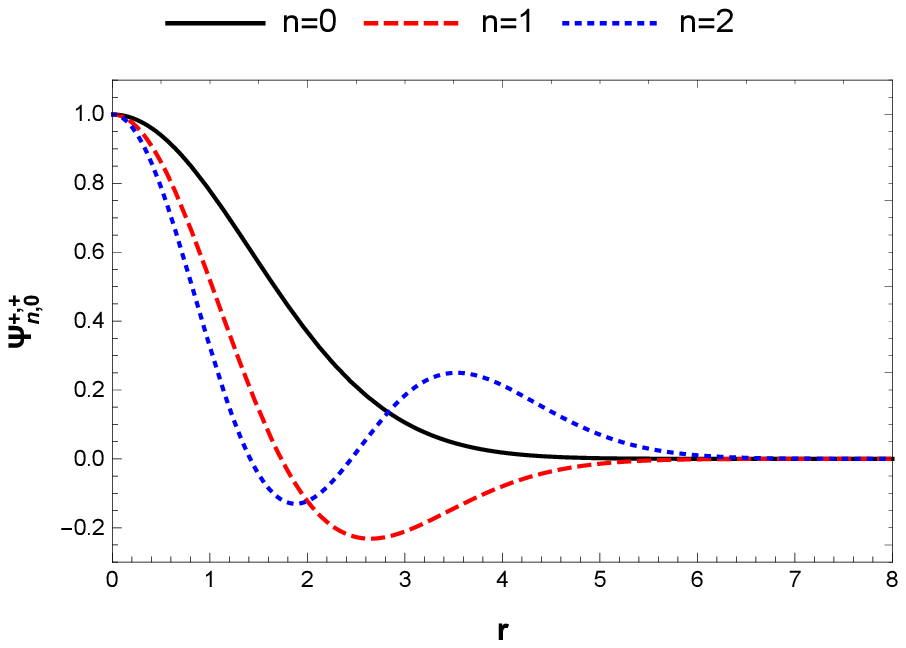},%
\includegraphics{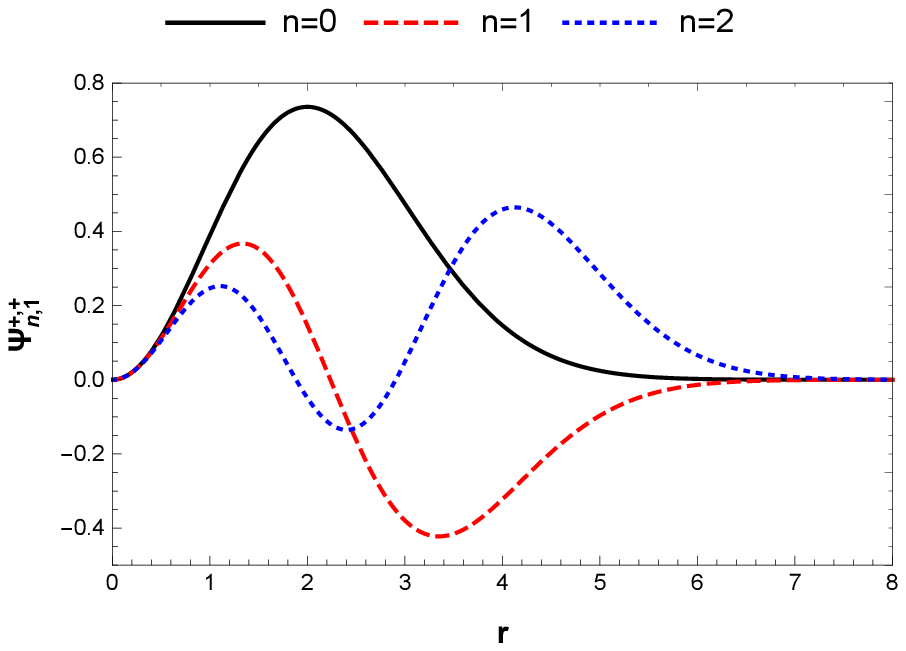}}
\caption{Unnormalized radial Dunkl-wave functions  of $\ell=0$ and $\ell=1$ cases, respectively. }\label{Fig2}
\end{figure*}

The result of this case can easily can be obtained by $\mu_1\rightarrow\mu_2$ and $\mu_2\rightarrow\mu_1$ out of the previous subsection. By doing so, we obtain the radial equation 
\begin{equation}
\left[ \xi \frac{d^{2}}{d\xi ^{2}}+\left( 1+\mu _{1}+\mu _{2}\right) \frac{d%
}{d\xi }-\frac{\xi }{4}-\frac{\lambda _{-}^{2}-4\mu _{1}\mu _{2}}{4\xi }+%
\frac{1+\mu _{2}-\mu _{1}-\lambda _{-}}{2}+\frac{l_{B}^{2}E^{2}}{2\hbar
^{2}v_{F}^{2}}\right] \Psi ^{-,+}=0,  \label{M}
\end{equation}%
and its eigensolution
\begin{equation}
\Psi ^{-,+}=\xi ^{\ell }e^{-\frac{\xi }{2}}\mathbf{F}\left( -n,1+2\ell+\mu
_{1}+\mu _{2} ,\xi \right),
\end{equation}
with the energy spectrum function
\begin{equation}
E_{n,\ell }^{-,+}=\pm \frac{\hbar v_{F}}{l_{B}}\sqrt{2\left(n+\ell +\mu _{1} \pm
\sqrt{\left( \ell +\mu _{1}\right) \left( \ell +\mu _{2}\right) }\right)}. \label{Enmp}
\end{equation} 
Similar to the previous case discussion, for ($\mu _{1}\rightarrow 0$ and $\mu _{2}\rightarrow 0$) we arrive at \eqref{firstsol}, while for 
($\mu _{1}+\mu_2\rightarrow 0$), we find
\begin{eqnarray}
E_{n,\ell}&=&\pm \frac{\hbar v_{F}}{l_{B}}\sqrt{2\left(n+\ell -\mu _{2} \pm \sqrt{ \ell^2 -\mu _{2}^2 }\right)}. \label{Enpm3}
\end{eqnarray}
Before we examine the thermal properties, we depict the Dunkl-energy spectra of four cases by taking $\hbar= v_{F}=l_{B}=1$ and by assigning two arbitrary values to the Wigner parameters, $\mu_1=0.75$ and $\mu_1=1.25$. In the first column of Fig. \ref{Fig1}, we demonstrate the positive signed expressions of Eqs. \eqref{Enpp} and \eqref{Enmm} for $\ell=1$. We observe that, alike the $\ell=0$ case, the spectra differs from the ordinary spectrum given in  Eq. \eqref{ordEnspec}. Similarly, in the second column of Fig. \ref{Fig1}, we present the energy spectra plots of $\ell= 1/2$ cases. Then, we show the ground state and first two excited states of $s_1=s_2=+1$ cases for two different values of $\ell$ in Fig. \ref{Fig2}. We see that the unnormalized wave function plots confirm the accuracy of our results. 

\section{Thermal Properties}
In this section, our goal is to determine thermal properties of Dunkl-graphene system. To this end, we have to derive the fundamental tool of statistical mechanics, namely the partition function of the model.

\subsection{Partition function}
\begin{figure*}[htb!]
\resizebox{\linewidth}{!}{\includegraphics{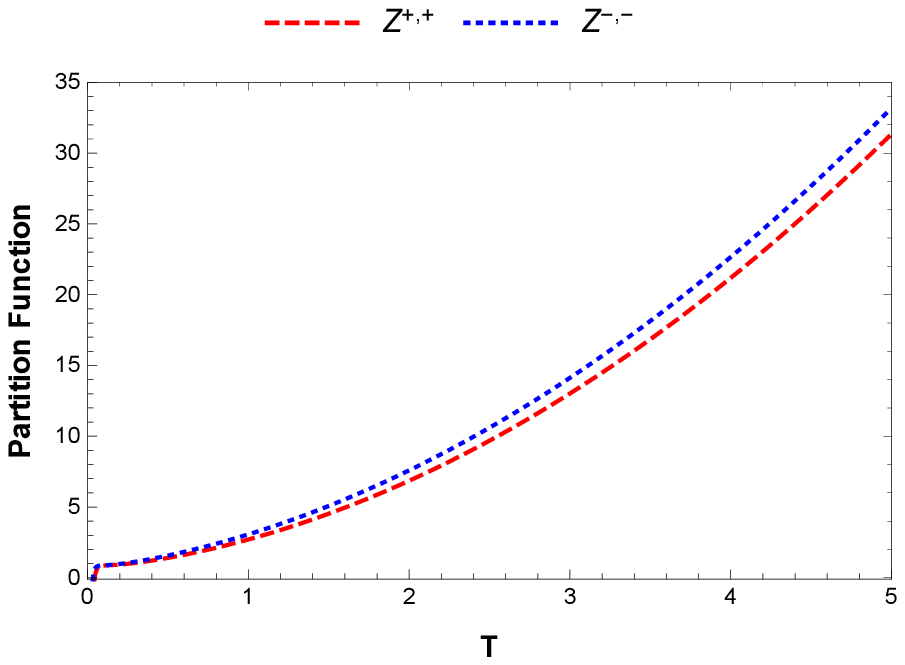},%
\includegraphics{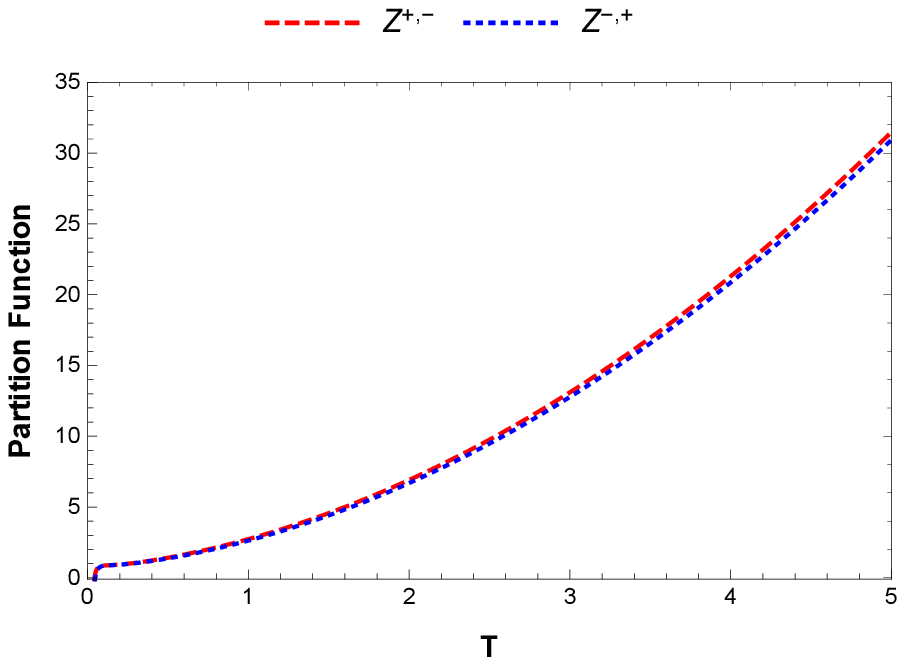}}
\caption{Dunkl-partition functions of different ensembles.}\label{Fig3}
\end{figure*}

Using the Dunkl-energy spectrum of graphene system, one can define the canonical partition function for fixed value of quantum number $\ell $ with a sum over all states $n$.
\begin{equation}
Z^{s_{1},s_{2}}=\sum_{n=0}e^{-\beta \left(
E_{n}^{s_{1},s_{2}}-E_{0}^{s_{1},s_{2}}\right) }.
\end{equation}%
Here, $\beta \equiv \frac{1}{K_{B}T}$, $K_{B}$ and $T$ are the Boltzmann constant  and the thermodynamic temperature, respectively. We take the energy eigenvalue in the following form
\begin{equation}
E_{n}^{s_{1},s_{2}}=\pm \frac{\hbar v_{F}}{l_{B}}\sqrt{2n+\varpi
_{s_{1},s_{2}}},
\end{equation}%
where%
\begin{equation}
\left \{ 
\begin{array}{l}
\varpi _{+,+}=2\left(\ell +\sqrt{\ell
\left( \ell +\mu _{1}+\mu _{2}\right) }\right), \\ 
\varpi _{-,-}=2\left( \ell +\mu
_{1}+\mu _{2} +\sqrt{\ell \left( \ell +\mu _{1}+\mu _{2}\right) }\right), \\ 
\varpi _{+,-}=2\left(\ell +\mu _{2}+\sqrt{\left( \ell +\mu _{1}\right) \left( \ell +\mu _{2}\right) }\right), \\ 
\varpi _{-,+}=2\left(\ell +\mu _{1}+\sqrt{\left( \ell +\mu _{1}\right) \left( \ell +\mu _{2}\right) }\right).%
\end{array}%
\right. 
\end{equation}
Then, we perform the computation of the summation over $n$ with the help of the Euler-Maclaurin formula%
\begin{equation}
\sum_{n=0}F\left( n\right) =\frac{1}{2}F\left( 0\right) +\int_{0}^{+\infty
}F\left( x\right) dx-\sum_{p=1}\frac{B_{2p}}{\left( 2p\right) !}F^{\left(
2p-1\right) }\left( 0\right).
\end{equation}
Here, $B_{2p}$ represents the Bernoulli numbers and $F^{\left( 2p-1\right) }\left(
0\right) $ denotes the derivative of order $2p-1$ of the function $F\left(x\right) $ at $x=0$. 
We obtain the Dunkl-partition function of different parity sectors as%
\begin{equation}
Z^{s_{1},s_{2}}\simeq \frac{1}{2}+\frac{l_{B}^{2}}{\hbar ^{2}\beta ^{2}v_{F}^{2}}+%
\frac{l_{B}}{\hbar \beta v_{F}}\sqrt{\varpi _{s_{1},s_{2}}}+\frac{1}{12}%
\frac{\beta \hbar v_{F}}{l_{B}\sqrt{\varpi _{s_{1},s_{2}}}}-\frac{1}{240}%
\frac{\hbar \beta v_{F}}{l_{B}\varpi _{s_{1},s_{2}}^{5/2}}-\frac{1}{240}%
\frac{\hbar ^{2}\beta ^{2}v_{F}^{2}}{l_{B}^{2}\varpi _{s_{1},s_{2}}^{2}}-%
\frac{1}{720}\frac{\hbar ^{3}\beta ^{3}v_{F}^{3}}{l_{B}^{3}\varpi
_{s_{1},s_{2}}^{3/2}}.
\end{equation}
In Fig. \ref{Fig3}, we compare the partition functions of four different ensembles. We would like to point out that when depicting the diagrams in the remainder of the manuscript we take the Boltzmann constant as one.

\subsection{Thermodynamics functions}
\begin{figure*}[htb!]
\resizebox{\linewidth}{!}{\includegraphics{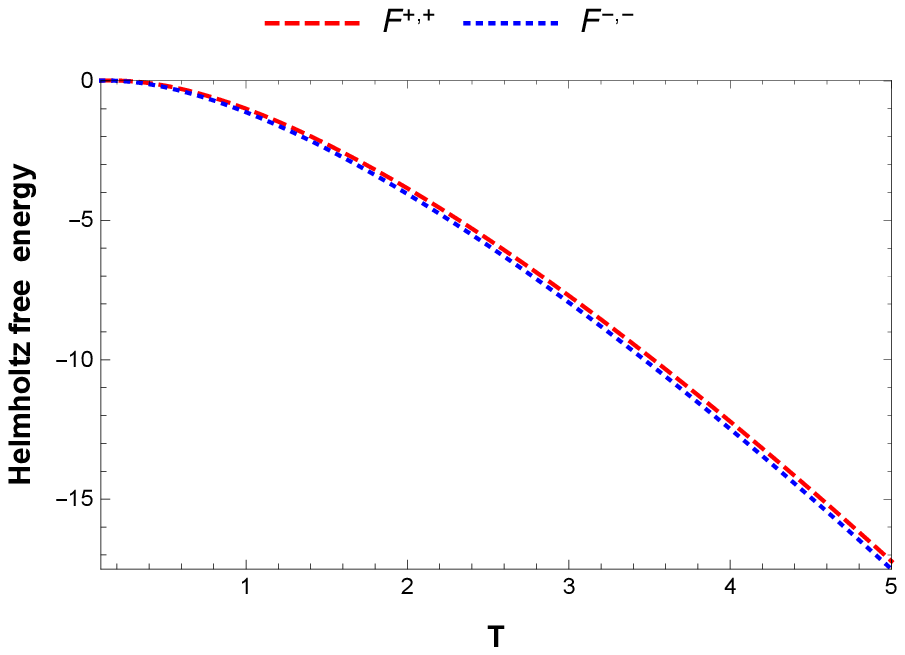},%
\includegraphics{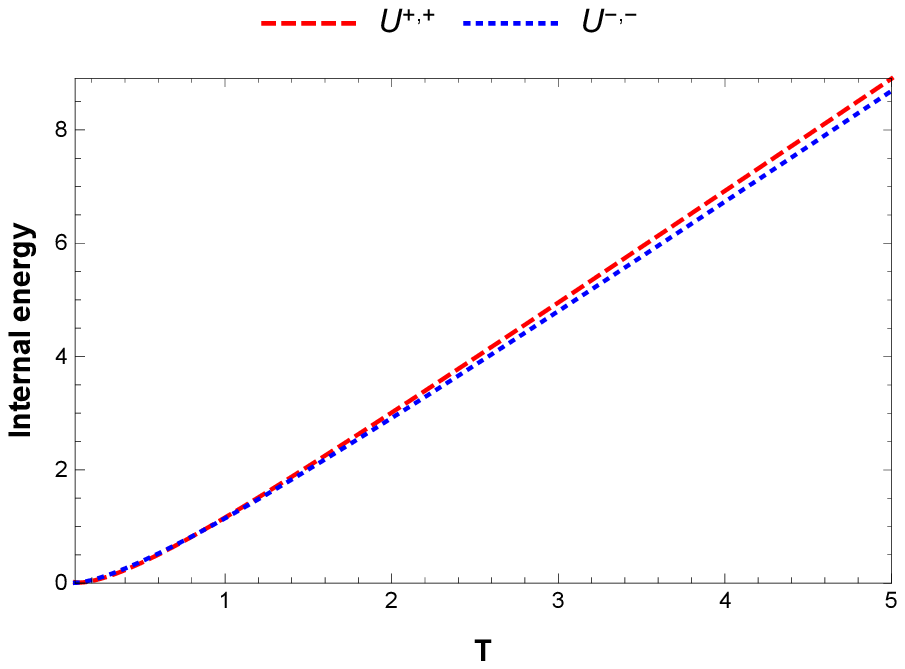}}\\
\resizebox{\linewidth}{!}{\includegraphics{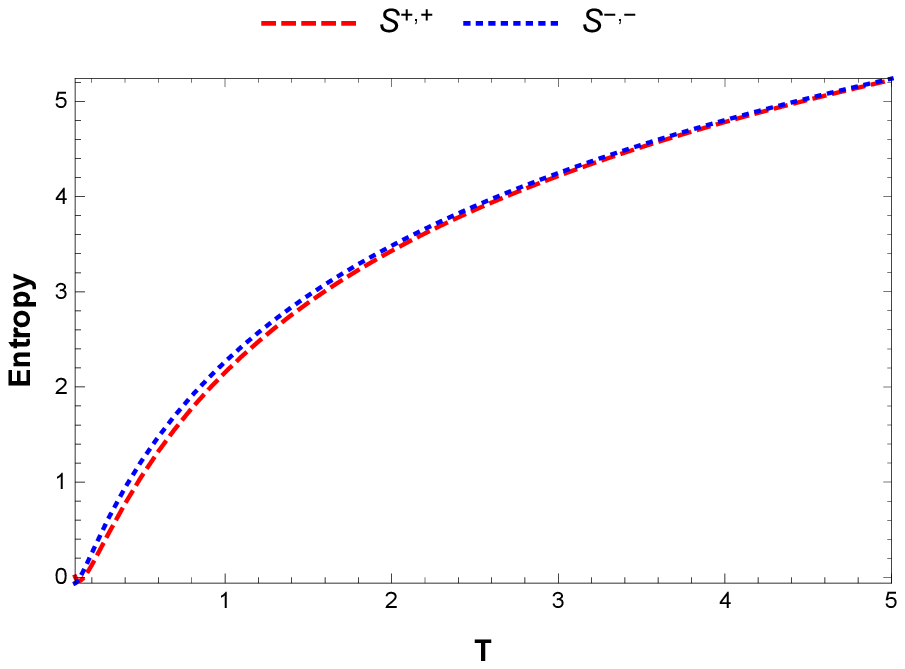},%
\includegraphics{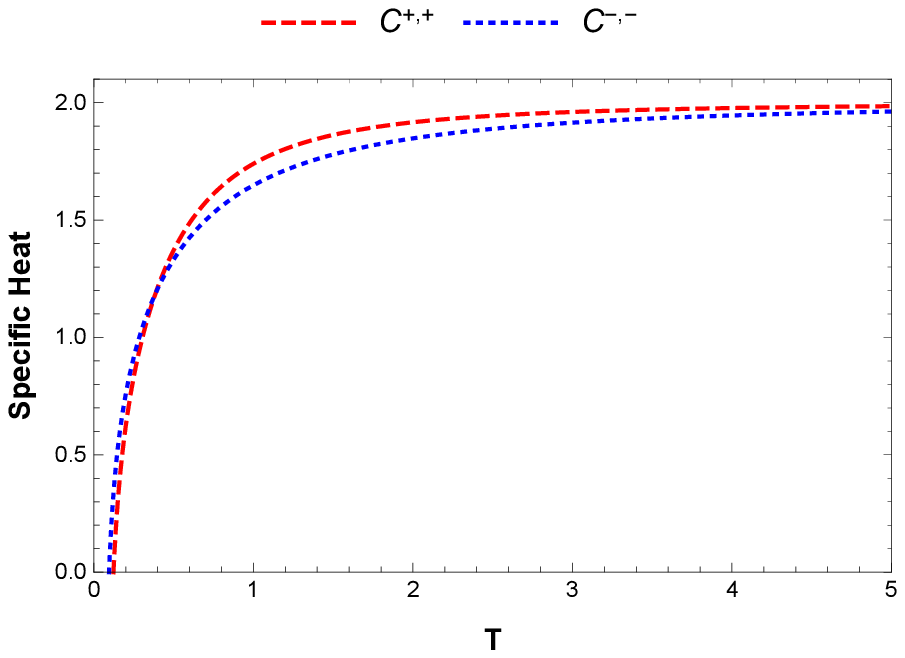}}
\caption{Dunkl-Thermodynamic functions versus temperature.}\label{Fig4}
\end{figure*}

Next, we use the partition function to obtain the most important thermodynamic functions: Helmholtz free energy, $F$, entropy, $S$, internal energy, $U$, and specific heat $C$, defined by the following
equations:%
\begin{equation}
F=-\frac{1}{\beta }\ln Z,\quad U=-\frac{\partial }{\partial \beta }\ln Z,\quad \frac{S}{K_{B}}=\beta ^{2}\frac{\partial F}{\partial \beta }, \quad \frac{C}{K_{B}}=-\beta ^{2}\frac{\partial U}{\partial \beta }.
\end{equation}
We find the Dunkl-Helmholtz free energy
\small
\begin{eqnarray}
F^{s_{1},s_{2}}\simeq -\frac{1}{\beta }\ln \left[ \frac{1}{2}+\frac{l_{B}^{2}}{%
\hbar ^{2}\beta ^{2}v_{F}^{2}}+\frac{l_{B}}{\hbar \beta v_{F}}\sqrt{\varpi
_{s_{1},s_{2}}}+\frac{1}{12}\frac{\beta \hbar v_{F}}{l_{B}\sqrt{\varpi
_{s_{1},s_{2}}}}-\frac{1}{240}\frac{\hbar \beta v_{F}}{l_{B}\varpi
_{s_{1},s_{2}}^{5/2}}-\frac{1}{240}\frac{\hbar ^{2}\beta ^{2}v_{F}^{2}}{%
l_{B}^{2}\varpi _{s_{1},s_{2}}^{2}}-\frac{1}{720}\frac{\hbar ^{3}\beta
^{3}v_{F}^{3}}{l_{B}^{3}\varpi _{s_{1},s_{2}}^{3/2}}\right], \label{DH} 
\end{eqnarray}
\normalsize
and the Dunkl-internal energy 
\begin{eqnarray}
U^{s_{1},s_{2}}\simeq \frac{\frac{2l_{B}^{2}}{\hbar ^{2}\beta ^{3}v_{F}^{2}}+\frac{%
l_{B}}{\hbar \beta ^{2}v_{F}}\sqrt{\varpi _{s_{1},s_{2}}}-\frac{1}{12}\frac{%
\hbar v_{F}}{l_{B}\sqrt{\varpi _{s_{1},s_{2}}}}+\frac{1}{240}\frac{\hbar
v_{F}}{l_{B}\varpi _{s_{1},s_{2}}^{5/2}}+\frac{1}{120}\frac{\hbar ^{2}\beta
v_{F}^{2}}{l_{B}^{2}\varpi _{s_{1},s_{2}}^{2}}+\frac{1}{240}\frac{\hbar
^{3}\beta ^{2}v_{F}^{3}}{l_{B}^{3}\varpi _{s_{1},s_{2}}^{3/2}}}{\frac{1}{2}+%
\frac{l_{B}^{2}}{\hbar ^{2}\beta ^{2}v_{F}^{2}}+\frac{l_{B}}{\hbar \beta
v_{F}}\sqrt{\varpi _{s_{1},s_{2}}}+\frac{1}{12}\frac{\beta \hbar v_{F}}{l_{B}%
\sqrt{\varpi _{s_{1},s_{2}}}}-\frac{1}{240}\frac{\hbar \beta v_{F}}{%
l_{B}\varpi _{s_{1},s_{2}}^{5/2}}-\frac{1}{240}\frac{\hbar ^{2}\beta
^{2}v_{F}^{2}}{l_{B}^{2}\varpi _{s_{1},s_{2}}^{2}}-\frac{1}{720}\frac{\hbar
^{3}\beta ^{3}v_{F}^{3}}{l_{B}^{3}\varpi _{s_{1},s_{2}}^{3/2}}}, \label{DU}
\end{eqnarray}%
and the reduced Dunkl-entropy
\begin{eqnarray}
\frac{S^{s_{1},s_{2}}}{K_{B}}&\simeq &\ln \left[\frac{1}{2}+\frac{l_{B}^{2}}{\hbar ^{2}\beta ^{2}v_{F}^{2}}+%
\frac{l_{B}}{\hbar \beta v_{F}}\sqrt{\varpi _{s_{1},s_{2}}}+\frac{1}{12}%
\frac{\beta \hbar v_{F}}{l_{B}\sqrt{\varpi _{s_{1},s_{2}}}}-\frac{1}{240}%
\frac{\hbar \beta v_{F}}{l_{B}\varpi _{s_{1},s_{2}}^{5/2}}-\frac{1}{240}%
\frac{\hbar ^{2}\beta ^{2}v_{F}^{2}}{l_{B}^{2}\varpi _{s_{1},s_{2}}^{2}}-%
\frac{1}{720}\frac{\hbar ^{3}\beta ^{3}v_{F}^{3}}{l_{B}^{3}\varpi
_{s_{1},s_{2}}^{3/2}}\right] \nonumber \\
&+& \beta \left[\frac{\frac{2l_{B}^{2}}{\hbar ^{2}\beta ^{3}v_{F}^{2}}+\frac{%
l_{B}}{\hbar \beta ^{2}v_{F}}\sqrt{\varpi _{s_{1},s_{2}}}-\frac{1}{12}\frac{%
\hbar v_{F}}{l_{B}\sqrt{\varpi _{s_{1},s_{2}}}}+\frac{1}{240}\frac{\hbar
v_{F}}{l_{B}\varpi _{s_{1},s_{2}}^{5/2}}+\frac{1}{120}\frac{\hbar ^{2}\beta
v_{F}^{2}}{l_{B}^{2}\varpi _{s_{1},s_{2}}^{2}}+\frac{1}{240}\frac{\hbar
^{3}\beta ^{2}v_{F}^{3}}{l_{B}^{3}\varpi _{s_{1},s_{2}}^{3/2}}}{\frac{1}{2}+%
\frac{l_{B}^{2}}{\hbar ^{2}\beta ^{2}v_{F}^{2}}+\frac{l_{B}}{\hbar \beta
v_{F}}\sqrt{\varpi _{s_{1},s_{2}}}+\frac{1}{12}\frac{\beta \hbar v_{F}}{l_{B}%
\sqrt{\varpi _{s_{1},s_{2}}}}-\frac{1}{240}\frac{\hbar \beta v_{F}}{%
l_{B}\varpi _{s_{1},s_{2}}^{5/2}}-\frac{1}{240}\frac{\hbar ^{2}\beta
^{2}v_{F}^{2}}{l_{B}^{2}\varpi _{s_{1},s_{2}}^{2}}-\frac{1}{720}\frac{\hbar
^{3}\beta ^{3}v_{F}^{3}}{l_{B}^{3}\varpi _{s_{1},s_{2}}^{3/2}}}\right], \label{DS}
\end{eqnarray}
and the reduced Dunkl-specific heat
\begin{eqnarray}
\frac{C^{s_{1},s_{2}}}{K_{B}} &\simeq& \frac{\frac{6l_{B}^{2}}{\hbar ^{2}\beta ^{2}v_{F}^{2}}+\frac{2l_{B}\sqrt{%
\varpi _{s_{1},s_{2}}}}{\hbar \beta v_{F}}-\frac{1}{120}\frac{\hbar
^{2}\beta ^{2}v_{F}^{2}}{l_{B}^{2}\varpi _{s_{1},s_{2}}^{2}}-\frac{1}{120}%
\frac{\hbar ^{3}\beta ^{3}v_{F}^{3}}{l_{B}^{3}\varpi _{s_{1},s_{2}}^{3/2}}}{%
\frac{1}{2}+\frac{l_{B}^{2}}{\hbar ^{2}\beta ^{2}v_{F}^{2}}+\frac{l_{B}}{%
\hbar \beta v_{F}}\sqrt{\varpi _{s_{1},s_{2}}}+\frac{1}{12}\frac{\beta \hbar
v_{F}}{l_{B}\sqrt{\varpi _{s_{1},s_{2}}}}-\frac{1}{240}\frac{\hbar \beta
v_{F}}{l_{B}\varpi _{s_{1},s_{2}}^{5/2}}-\frac{1}{240}\frac{\hbar ^{2}\beta
^{2}v_{F}^{2}}{l_{B}^{2}\varpi _{s_{1},s_{2}}^{2}}-\frac{1}{720}\frac{\hbar
^{3}\beta ^{3}v_{F}^{3}}{l_{B}^{3}\varpi _{s_{1},s_{2}}^{3/2}}} \nonumber \\
&-&\beta ^{2}\left( \frac{\frac{2l_{B}^{2}}{\hbar ^{2}\beta ^{3}v_{F}^{2}}+%
\frac{l_{B}\sqrt{\varpi _{s_{1},s_{2}}}}{\hbar \beta ^{2}v_{F}}-\frac{1}{12}%
\frac{\hbar v_{F}}{l_{B}\sqrt{\varpi _{s_{1},s_{2}}}}+\frac{1}{240}\frac{%
\hbar v_{F}}{l_{B}\varpi _{s_{1},s_{2}}^{5/2}}+\frac{1}{120}\frac{\hbar
^{2}\beta v_{F}^{2}}{l_{B}^{2}\varpi _{s_{1},s_{2}}^{2}}+\frac{1}{240}\frac{%
\hbar ^{3}\beta ^{2}v_{F}^{3}}{l_{B}^{3}\varpi _{s_{1},s_{2}}^{3/2}}}{\frac{1%
}{2}+\frac{l_{B}^{2}}{\hbar ^{2}\beta ^{2}v_{F}^{2}}+\frac{l_{B}}{\hbar
\beta v_{F}}\sqrt{\varpi _{s_{1},s_{2}}}+\frac{1}{12}\frac{\beta \hbar v_{F}%
}{l_{B}\sqrt{\varpi _{s_{1},s_{2}}}}-\frac{1}{240}\frac{\hbar \beta v_{F}}{%
l_{B}\varpi _{s_{1},s_{2}}^{5/2}}-\frac{1}{240}\frac{\hbar ^{2}\beta
^{2}v_{F}^{2}}{l_{B}^{2}\varpi _{s_{1},s_{2}}^{2}}-\frac{1}{720}\frac{\hbar
^{3}\beta ^{3}v_{F}^{3}}{l_{B}^{3}\varpi _{s_{1},s_{2}}^{3/2}}}\right) ^{2}. \label{DC}
\end{eqnarray}%
In the high temperature region, the dominant term of the Dunkl-partition function becomes independent of the Wigner parameters
\begin{eqnarray}
Z^{s_{1},s_{2}}=\frac{l_{B}^{2} K_{B}^{2}}{\hbar ^{2} v_{F}^{2}} T ^2,
\end{eqnarray}
which leads to a constant value to the specific heat function
\begin{equation}
C^{s_{1},s_{2}}=2K_{B}.
\end{equation}
We note that these limits follow the Dulong-Petit law for an
ultra-relativistic ideal gas. Finally, we depict and compare the derived Dunkl-thermodynamic functions versus temperature for positive and negative parity ensembles in Fig. \ref{Fig4}.

We  observe that parity effects appear on the thermal quantities at low temperatures. 

\newpage
\section{Conclusion}

In this manuscript, we solved the massless Dirac equation under the presence of a constant external magnetic field within the Dunkl-formalism. To this end after constructing the Dunkl-Dirac Hamilton operator, we employed polar coordinates to derive the energy spectra and the corresponding wave functions according to reflection symmetry concept. We demonstrated several ground and excited states as well as the energy spectra.  Then, we explored the Dunkl-Graphene thermal quantities by studying the most important thermodynamic functions. We observed that the parity effects show itself at low temperatures.

\section*{Acknowledgments}
One of the authors of this manuscript, BCL, is supported by the Internal  Project,  [2022/2218],  of  Excellent  Research  of  the  Faculty  of  Science  of Hradec Kr\'alov\'e University.

\section*{Data Availability Statements}
The authors declare that the data supporting the findings of this study are available within the article.

\end{document}